\documentclass[12pt]{article}

\usepackage{amssymb}
\usepackage{amsmath}

\usepackage{epsfig}

\begin{document}

\title{\bf Multiple hadron production in $\mathbf{e^+e^-}$ annihilation
induced by heavy primary quarks. New analysis}

\author{A.V. Kisselev\thanks{E-mail: alexandre.kisselev@ihep.ru} \
and V.A. Petrov\thanks{E-mail: vladimir.petrov@ihep.ru} \\
\small Institute for High Energy Physics, 142281 Protvino, Russia}

\date{}

\maketitle

\thispagestyle{empty}

\bigskip

\begin{abstract}
In this paper we present an analysis of the multiple hadron
production induced by primary heavy quarks in $e^+e^-$
annihilation with the account of most complete and corrected
experimental data. In the framework of perturbative QCD, new
theoretical bounds on the asymptotically constant differences of
the multiplicities in processes with light and heavy quarks are
given.
\end{abstract}



\section{Introduction}

As it is already known in classical theory, the more is the mass
of a charged particle, the less intensive is the radiation from
it. In quantum field theory, in particular, QCD, this circumstance
leads to a number of impressive effects caused by heavy quarks,
as, for instance, the effect of leading mesons which contain $c$-
and $b$-quarks in $e^+e^-$ annihilation~\cite{Kartvelishvili:78}.
If we assume that the ``radiation'' in QCD is the radiation of
gluons, which ``materialize'' afterwards in form of detected
hadrons, it is natural to expect, that in events induced by heavy
quarks, the hadron multiplicity (except for decay products of the
leading quarks) is smaller that in analogous events triggered by
light quarks. So natural may be an expectation that the difference
between these multiplicities has to disappear at high enough
energies.

However, it is not always the case. For instance, even in QED, the
difference between cross sections (times the c.m. energy squared)
of fermion-antifermion pairs of different masses produced by two
photons, does not disappear with increase of energy, but tends to
a constant which depends on the masses of the final
fermions~\cite{QED}. The measurements of hadron multiplicities in
$e^+e^-$ annihilation associated with the primary quarks of
definite flavors (in practice, $u$, $d$ and $s$ quarks are assumed
to be massless) were carried out at lepton colliders with the
collision energy of 29 GeV and more, in particular, at the SLC,
KEK, LEPI and LEPII.

In view of a great interest caused by these experiments (see
below), and due to the presence of competing theoretical
predictions, there is no doubts that similar measurements will be
also done at future linear colliders, such as CLIC or
ILC~\cite{LC}. The first attempt to look for such effects was done
in the framework of so-called ``na\"{\i}ve model''. The essence of
this model lies in the fact that an universal mechanism of
multiple production of hadrons from some gluon system is adopted,
which is insensitive to the quark flavors, while all the
difference between processes induced by distinct quark-antiquark
pairs arises due to a difference in energy available for the
hadron production.

Later on, another approach was accepted appealing directly to the
calculational kit  of perturbative QCD. However, the estimates
obtained this time overestimated the data significantly. For all
that, there was a clear indication of the asymptotic constancy of
the multiplicity differences from the heavy and light quarks. Not
long after, the same quantity, the multiplicity difference in the
events with the light and heavy quarks, was calculated more
accurately. The results appeared to be strikingly close to the
experimental data.

As was already said, the ``na\"{\i}ve model''~\cite{Rowson:85,
Kisselev:88} was the first attempt to take into account hadrons
produced in addition to decay products of the heavy
quark-antiquark pair in $e^+e^-$ annihilation. Later on, it was
argued~\cite{Schumm:92} that the difference between multiplicities
in events induced by heavy ($Q = c, \, b$) and light ($l = u, \,
d, \, s$) quarks,
\begin{equation}\label{delta_Ql}
\delta_{Ql} = N_{Q\bar{Q}}(Q^2) - N_{l\bar{l}}(Q^2),
\end{equation}
tends to a constant value with increase of energy $Q =
\sqrt{q^2}$:
\begin{equation}\label{MLLA_prediction}
\delta_{Ql} \rightarrow \delta_{Ql}^{\mathrm{MLLA}} = 2n_Q -
N_{l\bar{l}}(m_Q^2 \mathrm{e}).
\end{equation}
Here (and below) it is assumed that we deal with \emph{mean}
multiplicities of \emph{charged} hadrons, and ``$\mathrm{e}$'' is
the base of the natural logarithm ($\ln \mathrm{e} = 1$).

The comparison with the data has shown that the ``na\"{\i}ve
model'' describes the data on $\delta_{bl}$ up to $Q = 58$
GeV~\cite{Rowson:85, b_events_low_energies_1,
b_c_events_low_energies, b_events_low_energies_2}, but
underestimates the LEP and SLAC
data~\cite{b_events_high_energies_1, b_c_events_high_energies,
b_events_high_energies_2}. As for the formula which was obtained
at the basis of the co-called ``modified logarithmic approximation
(MLLA)''~\eqref{MLLA_prediction}, it has significantly
overestimated both low and high-energy data on $\delta_{bl}$.

Detailed QCD calculations of the difference between associated
multiplicities of charged hadron in $e^+e^-$ annihilation were
made in \cite{Petrov:95}. The QCD expressions for $\delta_{Ql}$
from Ref.~\cite{Petrov:95} appeared to be in a good agreement with
experimental measurements of associated hadron multiplicities in
$e^+e^-$ annihilation (see, for instance, \cite{Chrin:94,
Metzger:95}).

As we will see below, it is the hadron multiplicity in the light
quark events that enables one to calculate the multiplicity
differences $\delta_{Ql}$. Recently, the data on average charged
multiplicities in $l \bar{l}$ events at different energies
corrected for detector effects as well as for initial state
radiation were presented in \cite{Dokshitzer:06}. The corrected
multiplicity differences averaged over all presently published
results can be found in Ref.~\cite{Dokshitzer:06}:
\begin{align}
\delta_{bl}^{\rm exp} &= 3.12 \pm 0.14 \;, \label{delta_bl_exp}
\\
\delta_{cl}^{\rm exp} &= 1.0 \pm 0.4 \;. \label{delta_cl_exp}
\end{align}

Because of the appearance of these corrected experimental data, a
natural necessity has arisen to reconsider our predictions for
$\delta_{bl}$  with the account of the data on the hadron
multiplicity in the light quark events $N_{l\bar{l}}(Q^2)$ as
well.

In Section~\ref{sec:e+e-_multiplicity} we define all the relevant
quantities, since rigorous definitions are not usually presented
by other authors. The analytical formula for the hadron
multiplicity in $e^+e^-$ annihilation is obtained. The QCD
expression for the multiplicity difference is derived in the next
section. In Section~\ref{sec:bounds} the upper and lower bounds on
$\delta_{bl}$ are calculated. In Appendix~A the detailed
derivation of the evolution equation for the multiplicity in the
gluon jet is given. In Appendix~B we discuss the connection
between our approach and the scheme which uses the concept of the
Altarelli-Parisi decay functions. The problems related to the
gauge invariance guarantee which appear in perturbative
calculations of the ``light multiplicity'' are considered in
Appendix~C.



\section{Hadron multiplicities in $\mathbf{e^+e^-}$ annihilation}
\label{sec:e+e-_multiplicity}

The average multiplicity of hadrons in a $q\bar{q}$ event in the
process of $e^+e^-$ annihilation is of the form
\begin{equation}\label{hadron_mult}
N_{q\bar{q}}^h(Q^2) = 2 n_{q} + \!\int \! \frac{d^4k}{(2\pi)^4} \,
\Pi_{\mu \nu}^{a b}(q,k) \, d^{\, \mu \alpha}_{a a'}(k) \, d^{\,
\nu \beta}_{b b'}(k) \, n^{a' b'}_{\alpha \beta}(k) \;,
\end{equation}
where $d^{\, \mu \nu}_{a b}(k)$ is the propagator of the gluon
with momentum $k$. Here and below $(a, b)$ and $(a', b')$ denote
color indices.

The subscript $q$ denotes the type of primary quarks. In what
follows, the notation $q=Q$ (\emph{heavy quark}) will mean charm
or beauty quark, while the notation $q=l$ (\emph{light quark})
designates a pair of $u,\ d$ or $s$-quarks  which are assumed to
be massless. In particular, $N_{l\bar{l}}(Q^2)$ means the
multiplicity of hadrons in light quark events, while
$N_{Q\bar{Q}}(Q^2)$ is the multiplicity of hadrons in events when
the process is induced by the heavy quark and antiquark of the
type $Q$.

\begin{figure}[htb]
\begin{center}
\epsfysize=5cm \epsffile{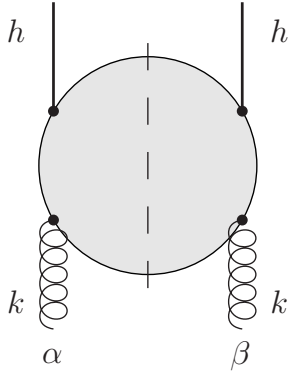}
\end{center}
\caption{The diagram which describes the average multiplicity of
hadrons with the 4-momentum $h$ (solid lines) in the gluon jet
with the virtuality $k^2$ (winding lines).}
\label{fig:frag}
\end{figure}

The first term in the r.h.s. of Eq.~\eqref{hadron_mult}, $2n_q$,
is the multiplicity from the fragmentation of the leading quark
(antiquark). It is taken from the analysis of data ($2n_c = 5.2$,
$2n_b = 11.0$~\cite{Schumm:92}, and $2n_l = 2.4$~\cite{Chrin:94}).
The quantity $n^{a' b'}_{\alpha \beta}(k)$ in the integrand of
Eq.~\eqref{hadron_mult} is given by the diagram in
Fig.~\ref{fig:frag} in which both the integration in the momentum
of the final hadron and averaging in its polarization are assumed.
This diagram corresponds to the following analytical expression:
\begin{align}\label{gluon_frag_tensor}
n^{a' b'}_{\alpha \beta}(k) &= \int \!\! \frac{d^3h}{(2\pi)^3
2h_0} \iiint \! d^4x \, d^4y \, d^4z \, \mathrm{e}^{ikx - ih(y-z)}
\nonumber \\
& \times \langle 0| \Big( T I_{\alpha}^{a'}(x) J_h^{+}(y) \Big)
\Big( T I_{\beta}^{b'} (0) J_h(z) \Big)|0 \rangle
\nonumber \\
&= - i \! \iiint \! d^4x \, d^4y \, d^4z \, \mathrm{e}^{ikx}
D_h^{-}(y-z)
\nonumber \\
& \times \langle 0| \Big( T I_{\alpha}^{a'}(x) J_h^{+}(y) \Big)
\Big( T I_{\beta}^{b'} (0) J_h(z) \Big)|0 \rangle \;.
\end{align}
Here $J_h(x)$ is the source operator of hadron $h$ (spin indices
are omitted), and $I_{\alpha}^a(x)$ is the gluon (color)
current~\cite{Bogolyubov},
\begin{equation}\label{color_current}
I_{\mu}^a (x) = i \frac{\delta S}{\delta A^{\mu}_a(x)} S^{\ast}
\;,
\end{equation}
where $A_a^{\mu}(x)$ denotes the gluon field. $D_h^{-}(x)$ is the
Pauli-Jordan function ~\cite{Bogolyubov}:
\begin{equation}\label{D_func}
D_h^{-}(x) = i \, \langle 0|h(x) h(0)|0 \rangle \;,
\end{equation}
where $h(x)$ is an asymptotic hadronic field operator.%
\footnote{For half-spin hadrons, $D_h^{-}(x)$ is replaced by
$S_h^{-}(x) = (i \hat{\partial} + m_h) D_h^{-}(x)$.}

Since $k^{\alpha}I_{\alpha}^a(x) = 0$, and the final hadrons are
colorless particles, we get:
\begin{equation}\label{gluon_frag}
n^{a' b'}_{\alpha \beta}(k) = \left( -g_{\alpha \beta} \, k^2 +
k_{\alpha} k_{\beta} \right) \delta^{a' b'} n_g(k^2) \;,
\end{equation}
where dimensionless quantity $n_g(k^2)$ describes the average
multiplicity of hadrons in the gluon jet with the virtuality
$k^2$. It is, of course, gauge invariant, and depends only on the
virtuality $k^2$.

It is useful to introduce the average multiplicity from the gluon
jet whose \emph{virtuality $p^2$ varies up to} $k^2$:
\begin{equation}\label{N_g_vs_n_g}
N_g(k^2, Q_0^2) = \int^{k^2}_{Q_0^2} \!\! \frac{d p^2}{p^2} \,
n_g(p^2) \;.
\end{equation}
Very often $N_g(k^2)$ is erroneously called the average
multiplicity of the gluon jet with \emph{fixed virtuality} $k^2$.
This meaning should be addressed to $n_g(k^2)$ only.

The infrared cut-off  $Q_0$ which separates perturbative and
non-perturbative regions was introduced in \eqref{N_g_vs_n_g}. We
will use the ``conventional standard'' value $Q_0 = 1$ GeV for our
numerical estimates (see Section~\ref{sec:bounds}).

The quantity $n_g(k^2)$ cannot be calculated perturbatively. It is
usually assumed that the average hadron multiplicity is
proportional to $n_g(k^2, Q_0^2)$, i.e. the average multiplicity
of (off-shell) partons with the ``mass'' $Q_0$ (the so-called
local parton-hadron duality):
\begin{equation}\label{LPHD}
n_g(k^2) = n_g(k^2, Q_0^2) \, K(Q_0^2) \;,
\end{equation}
where $K(Q_0^2)$ is a phenomenological energy-independent factor.
The QCD evolution equations for both $n_g(k^2, Q_0^2)$ and
$N_g(k^2, Q_0^2)$ are derived in Appendix~A. Let us stress,
however, that the main results of the present paper (see Sections
\ref{sec:mult_difference}, \ref{sec:bounds}) \emph{do not depend
on explicit form} of the function $n_g(k^2)$.

In Eq.~\eqref{hadron_mult} the first factor of the integrand is
given by
\begin{equation}\label{convolution}
\Pi_{\mu \nu}^{ab}(q,k) = (-g^{\rho \sigma}) \Pi_{\rho \sigma; \mu
\nu}^{ab}(q,k) \;,
\end{equation}
where $\Pi_{\rho \sigma; \mu \nu}^{ab}(q,k)$ is the two-gluon
irreducible part of the relevant discontinuity of the
four-current correlation function%
\footnote{Note that $\Pi_{\rho \sigma; \mu \nu}^{ab}$ is
proportional to $\delta^{ab}$.}
\begin{align}\label{correlation_func}
\Pi_{\rho \sigma; \mu \nu}^{ab}(q,k) &= \iiint \! d^4x \, d^4y \,
d^4z \, \mathrm{e}^{iqx - ik(y-z)}
\nonumber \\
& \times \langle 0| \Big( T J_{\rho}^{\mathrm{em}}(x) I_{\mu}^a
(y) \Big) \Big( T J_{\sigma}^{\mathrm{em}}(0) I_{\nu}^b (z)\Big)|0
\rangle \;.
\end{align}
Here $J_{\rho}^{\mathrm{em}}(x)$ is the electromagnetic current.
The color current$I_{\mu}^a (x)$ was defined
above~\eqref{color_current}.

\begin{figure}[htb]
\begin{center}
\epsfysize=4.5cm \epsffile{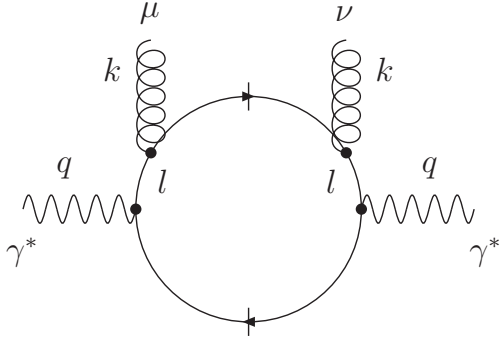}
\end{center}
\caption{The inclusive distribution of the massive gluon jet
(winding line) with the virtuality $k^2$. The wavy line is the
virtual photon with 4-momentum $q$. The solid lines are light
quarks. The cut lines correspond to on-shell quarks.}
\label{fig:ladder}
\end{figure}

\begin{figure}[htb]
\begin{center}
\epsfysize=4.5cm \epsffile{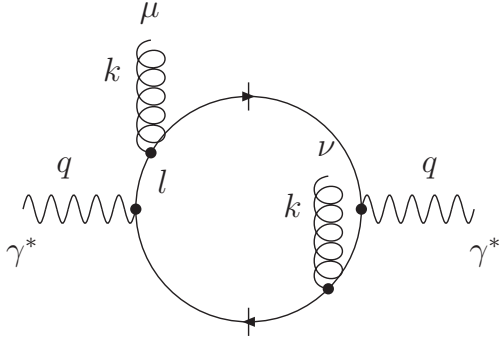}
\end{center}
\caption{The interference diagram which also contributes to the
inclusive distribution of gluon jets.}
\label{fig:cross}
\end{figure}

In the first order in the strong coupling constant, $\Pi_{\mu
\nu}^{ab}(q,k)$  is given by two diagrams presented in
Fig.~\ref{fig:ladder} and Fig.~\ref{fig:cross} normalized to the
total $e^+e^-$ rate.
This quantity is gauge invariant:
\begin{equation}\label{P_conserv}
k^{\mu} \Pi_{\mu \nu}^{ab}(q,k) = 0 \;.
\end{equation}

Let us define
\begin{align}\label{E}
C_F \, \frac{\alpha_s(k^2)}{\pi k^2} E(Q^2, k^2) &=
\frac{1}{(2\pi)^3 Q^2} \Big( \!\! - k^2 \frac{\partial}{\partial
k^2} \Big) \! \int \! d(qk) \sqrt{(qk)^2 - Q^2 k^2}
\nonumber \\
& \times (-g^{\mu \nu}) \delta_{ab} \Pi_{\mu \nu}^{ab}(Q^2, k^2,
qk) \;,
\end{align}
where $C_F = (N_c^2 - 1)/2 N_c$, and $N_c=3$ is the number of
colors. Then the average multiplicity in $e^+e^-$ annihilation
\eqref{hadron_mult} looks like
\begin{equation}\label{qq_mult}
N_{q\bar{q}}^h(Q^2, Q_0^2) = 2 n_{q} + C_F \!\int_{Q_0^2}^{Q^2} \!
\frac{dk^2}{k^2} \,\frac{\alpha_s(k^2)}{\pi} \, E(Q^2, k^2) \,
N_g(k^2) \;.
\end{equation}
The term $E(Q^2, k^2)$  is the inclusive spectrum of the gluon
jet with the virtuality up to $k^2$ emitted by primary quarks.%
\footnote{It was explained in detail in Ref.~\cite{Petrov:95} that
one should not consider this mechanism of hadron production via
gluon jets as due to ``a single cascading gluon'', as some authors
believe~\cite{Dokshitzer:06}. That $E$ is an inclusive spectrum of
the gluon jets is seen, e.g., from the fact that the average
number of gluon jets $\int (dk^2\!/k^2) \,E(Q^2, k^2) > 1$.}

For the light quark case, the explicit form of $ E(Q^2, k^2)$ was
obtained in Ref.~\cite{Petrov:95}. In terms of the variable
\begin{equation}\label{sigma_vs_eta}
\sigma = \frac{k^2}{Q^2} \;,
\end{equation}
we get:
\begin{eqnarray}\label{E_expression}
E(\sigma)&=& (1 + 2\sigma + 2\sigma^2) \ln \frac{1}{\sigma} -
\frac{3 + 7\sigma}{2} (1 - \sigma) - \sigma (1 + \sigma) \left(
\ln \frac{1}{\sigma} \right)^2
\nonumber \\
&+& 4\sigma(1 + \sigma) I(\sigma) \;,
\end{eqnarray}
where
\begin{equation}\label{I}
I(\sigma) = \frac{\pi^2}{4} - \mathrm{\rm Li}_2(1 + \sigma) \;,
\end{equation}
and $\mathrm{\rm Li}_2(z)$ is the Euler dilogarithm.

Let us introduce new variables
\begin{equation}\label{eta}
\eta = \ln \frac{Q^2}{k^2}
\end{equation}
and
\begin{equation}\label{Y}
Y = \ln \frac{Q^2}{Q_0^2} \;,
\end{equation}
as well as the notation
\begin{equation}\label{N_g_reduced}
\hat{N}_g(k^2) = C_F \, \frac{ \alpha_s(k^2)}{\pi} \, N_g(k^2) \;.
\end{equation}
Then Eq.~\eqref{qq_mult} can be rewritten as
\begin{equation}\label{mult_Y}
N_{q\bar{q}}(Y) = 2\,n_{q} + \int\limits_0^Y \! d \eta \,
\hat{N}_g(Y - \eta) \, E(\eta) = 2 \, n_{q} +  N_q(Y) \;.
\end{equation}
The physical meaning of the function
\begin{equation}\label{cent_mult_Y}
N_q(Y) = \int\limits_0^Y \! d \eta \, \hat{N}_g(Y - \eta) \,
E(\eta) \;
\end{equation}
in Eq.~\eqref{mult_Y} is the following. It describes the average
number of hadrons produced from virtual gluon jets emitted by
primary quark (antiquark) of the type $q$. In other words, it is
the multiplicity in $q\bar{q}$ event except for the multiplicity
of the decay products of these quarks at the final stage of
hadronization (the first term in \eqref{mult_Y}).

\begin{figure}[ht]
\epsfysize=6cm \epsffile{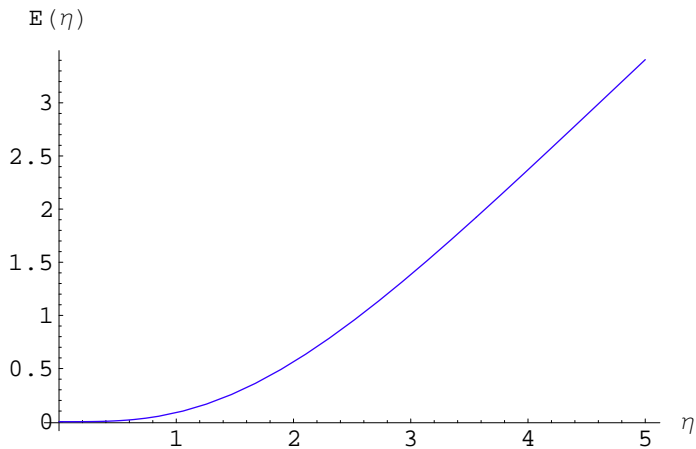} \caption{The function
$E(\eta)$.}
\label{fig:E}
\end{figure}

The function $E(\eta)$ is presented in Fig.~\ref{fig:E}. It has
the asymptotics
\begin{equation}\label{E_asym}
E(\eta)|_{\eta \rightarrow \, \infty} = E^{\mathrm{asym}}(\eta) =
\eta - \frac{3}{2} \;,
\end{equation}
The derivative of $E(\eta)$ is positive for all $\eta$, as one can
see in Fig.~\ref{fig:E}. It follows from the relation $\partial
N_{q\bar{q}}(Y)/\partial Y = \int_0^Y \! d \eta \, \hat{N}_g(\eta)
\, \partial E(Y - \eta)/\partial Y$ that the associative
multiplicity $N_{q\bar{q}}(Q)$ \eqref{mult_Y} is a monotone
increasing function of $Q$ for any positive function $N_g(k^2)$.

\section{Multiplicity difference in QCD}
\label{sec:mult_difference}

Now let us consider the difference between multiplicities in
events induced by the light and heavy quarks, $\delta_{Ql}$, which
is defined by Eq.~\eqref{delta_Ql}. The following representation
was found in Ref.~\cite{Petrov:95}:
\begin{equation}\label{delta_Ql_QCD}
\delta_{Ql}^{\mathrm{QCD}} = 2(n_Q - n_l) - \Delta N_Q(Y_m) \;.
\end{equation}
Here the new notation is introduced,
\begin{equation}\label{delta_N}
\Delta N_Q(Y_m) = N_q - N_Q = \int\limits_{-\infty}^{Y_m} \! dy \,
\hat{N}_g (Y_m - y) \, \Delta E_Q(y) \;,
\end{equation}
as well as variables
\begin{equation}\label{y}
y = \ln \frac{m_Q^2}{k^2} \;,
\end{equation}
and
\begin{equation}\label{Ym}
Y_m = \ln \frac{m_Q^2}{Q_0^2} \;.
\end{equation}

An important result which was obtained in Ref.~\cite{Petrov:95} is
that the function
\begin{equation}\label{delta_E_def}
\Delta E_Q =  E - E_Q
\end{equation}
depends only on the variable
\begin{equation}\label{ro_vs_y}
\rho = \exp (-y) \;,
\end{equation}
but not on the energy Q. The explicit form of $\Delta E_Q$ is the
following:
\begin{eqnarray}\label{Delta_E}
\Delta E_Q(\rho) &=& (1 - 3\rho + \frac{7}{2} \, \rho^2) \ln
\frac{1}{\rho} + \rho (7\rho - 20) \, J(\rho) + \frac{20}{\rho -
4} [1 - J(\rho)]
\nonumber \\
&+& 7\rho + \frac{9}{2} \;,
\end{eqnarray}
where
\begin{equation}\label{J}
J(\rho) =
  \begin{cases}
    \sqrt{\frac{\rho}{\rho - 4}}
    \ln \left( \frac{\sqrt{\rho} + \sqrt{\rho - 4}}{2}
    \right),
    & \rho > 4 \;, \cr
    \ 1 \;, & \rho = 4 \;, \cr
    \sqrt{\frac{\rho}{4 - \rho}}
    \arctan \left( \frac{\sqrt{4 - \rho}}{\rho} \right),
    & \rho < 4 \; .
  \end{cases}
\end{equation}

\begin{figure}[ht]
\epsfysize=6cm \epsffile{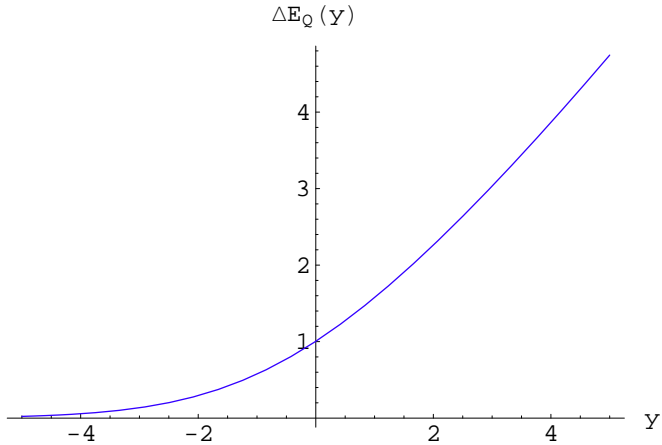} \caption{The function
$\Delta E_Q(y)$.} \label{fig:deltaE}
\end{figure}

Since $\Delta E_Q(y)$ has the asymptotics
\begin{equation}\label{Delta_E_asym}
\Delta E_Q(y) \Big|_{y \rightarrow -\infty} \simeq \frac{11}{3} \,
\exp(-|y|) \;,
\end{equation}
the integral in Eq.~\eqref{delta_N} converges rapidly at $y
\rightarrow -\infty$. The function $\Delta E_Q(y)$ is shown in
Fig.~\ref{fig:deltaE}. We find that
\begin{equation}\label{delta_E_asym}
\Delta E_Q(y)|_{y \rightarrow \, \infty} = \Delta
E_Q^{\mathrm{asym}}(y) = y - \frac{3}{2} \;.
\end{equation}

Another important relation comes from \eqref{E_asym} and
\eqref{delta_E_asym}:
\begin{equation}\label{diff_E_asym_reg}
\Delta E_Q(y - 1) - E(y) \Big|_{y \rightarrow \, \infty}  \simeq
\frac{5}{2} \, \sqrt{\mathrm{e}} \, \ln 2 \, \exp(-y/2) \;.
\end{equation}
In other words,
\begin{equation}\label{diff_E_asym}
\Delta E_Q(y) \simeq E(y+1)
\end{equation}
\emph{at large} $y$.

If one puts $\Delta E_Q(y) = E(y+1)$, then (neglecting the
contribution from the region $y < -1$):
\begin{equation}\label{delta_N_asym}
\Delta N_Q = N_{l\bar{l}}(m_Q^2\mathrm{e}) - 2\,n_l \;.
\end{equation}
Correspondingly, the approximate expression for $\delta_{Ql}$ is
of the form:
\begin{equation}\label{delta_asym}
\delta_{Ql}^{\mathrm{appr}} = 2n_Q - N_{l\bar{l}}(m_Q^2
\mathrm{e}) = \delta_{Ql}^{\mathrm{MLLA}} \;,
\end{equation}
where $\delta_{Ql}^{\mathrm{MLLA}}$ is the MLLA prediction for the
multiplicity difference~\cite{Schumm:92}. We would remind that the
function $N_{l\bar{l}}(Q)$ describes the hadron multiplicity in
light quark events at the collision energy $Q$.

\begin{figure}[ht]
\epsfysize=6cm \epsffile{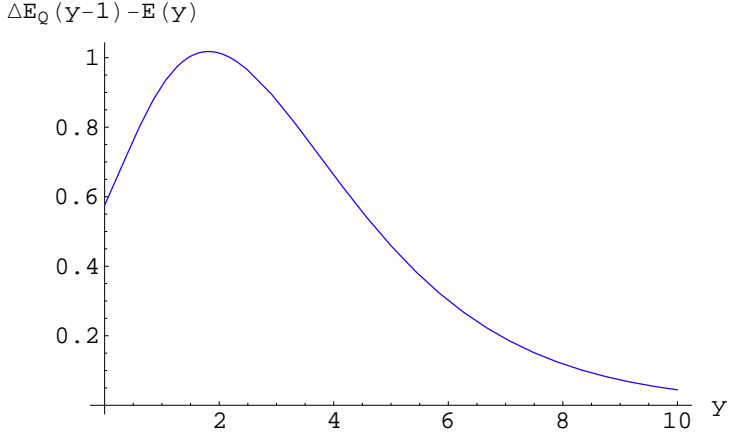} \caption{The
difference $\Delta E_Q(y - 1) - E(y)$ as a function of the
variable $y$.}
\label{fig:deltaE_vs_E}
\end{figure}

However, expression \eqref{diff_E_asym} is very far from the exact
one in the region $y < Y_m$,%
\footnote{For the beauty case, one has $Y_m \lesssim 3.2$.}
as it is clearly seen in Fig.~\ref{fig:deltaE_vs_E}. That is why,
there is a large difference between
$\delta_{Ql}^{\mathrm{MLLA}}$~\eqref{delta_asym} and QCD
expression $\delta_{Ql}^{\mathrm{QCD}}$~\eqref{delta_Ql_QCD}.

\section{Multiplicity difference: upper and lower bounds}
\label{sec:bounds}

In this section we will show that the difference between
$\delta_{Ql}^{\mathrm{MLLA}}$~\eqref{delta_asym} and
$\delta_{Ql}^{\mathrm{QCD}}$ is indeed numerically large, and also
obtain both upper and lower bounds on $\delta_{bl}$.

It is convenient to represent expression for $\Delta
N_Q$~\eqref{delta_N} in the form:
\begin{eqnarray}\label{Delta_N_expan}
\Delta N_Q(Y_m) &=& \int\limits_{0}^{Y_m + 1} \! dy \, \hat{N}_g
(Y_m + 1 - y) E(y)
\nonumber \\
 &+&  \int\limits_{-\infty}^{-1} \! dy \, \hat{N}_g (Y_m - y)
\, \Delta E_Q(y)
\nonumber \\
&+&  \int\limits_{0}^{Y_m + 1} \! dy \, \hat{N}_g (Y_m + 1 - y)
[\Delta E_Q(y - 1) - E(y)]
\nonumber \\
&\equiv& [N_{l\bar{l}}(m_Q^2 \mathrm{e}) - 2\,n_l ] + \delta
N_Q^{(1)}(Y_m ) + \delta N_Q^{(2)}(Y_m) \;,
\end{eqnarray}
that results in  the formula (see Eq.~\eqref{delta_Ql_QCD})
\begin{eqnarray}\label{delta_asym_corr}
\delta_{Ql}^{\mathrm{QCD}} &=& 2n_Q - N_{l\bar{l}}(m_Q^2
\mathrm{e})- \delta N_Q^{(1)}(Y_m) - \delta N_Q^{(2)}(Y_m)
\nonumber \\
&=& \delta_{Ql}^{(appr)} - \delta N_Q^{(1)}(Y_m ) - \delta
N_Q^{(2)}(Y_m)\;.
\end{eqnarray}
Here we have introduced notations
\begin{equation}\label{delta_N1_def}
N_Q^{(1)}(Y_m) = \int\limits_{-\infty}^{-1} \! dy \, \hat{N}_g
(Y_m - y) \, \Delta E_Q(y) \;,
\end{equation}
and
\begin{equation}\label{delta_N2_def}
N_Q^{(2)}(Y_m) = \int\limits_{0}^{Y_m + 1} \! dy \, \hat{N}_g (Y_m
+ 1 - y) [\Delta E_Q(y - 1) - E(y)] \;.
\end{equation}
Note that both $N_Q^{(2)}(Y_m)$ and $N_Q^{(2)}(Y_m)$ are positive
since $\Delta E_Q(y) > 0$ at all $y$ and $\Delta E_Q(y - 1) - E(y)
> 0$ at $y \geqslant 0$ (see Fig.~\ref{fig:deltaE} and
Fig.~\ref{fig:deltaE_vs_E}).

In order to exploit the corrected data on $N_{l\bar{l}}(Q^2)$ at
$Q =8 \mathrm{\ GeV}$,
\begin{equation}\label{N(mbe)_MLLA}
N_{l\bar{l}}(8.0 \mathrm{\ GeV}) = 6.70 \pm 0.34 \;,
\end{equation}
we assume that $m_b = 4.85$ GeV, that corresponds to $m_b
\sqrt{\mathrm{e}} = 8$ GeV.

The estimates show that the dominant correction to
$\delta_{Ql}^{\mathrm{QCD}}$ is $\delta N_Q^{(2)}$, not  $\delta
N_Q^{(1)}$. To calculate a lower bound for $\delta N_b^{(2)}$, we
will use the following inequality in the region $y \geqslant 0$:
\begin{equation}\label{deltaE_E}
\Delta E_Q(y) \geqslant E(y + \Delta y_Q) \;.
\end{equation}
The quantity $\Delta y_Q$ is a solution of the equation
\begin{equation}\label{delta_y}
\Delta E_Q(Y_m) = E(Y_m + \Delta y_Q) \;,
\end{equation}
where $Y_m$ is defined above~\eqref{Ym}. Then we get from
Eqs.~\eqref{delta_N2_def} and \eqref{deltaE_E}:
\begin{equation}\label{delta_N2_bound}
\delta N_Q^{(2)} \geqslant N_{l\bar{l}}(Y_m + \Delta y_Q) -
N_{l\bar{l}}(Y_m + 1) - \int\limits_{0}^{\Delta y_Q - 1} \! dy \,
\hat{N}_g (Y_m + \Delta y_Q - y) E(y) \;.
\end{equation}

For our further estimates, we need to know the hadron multiplicity
in the light quark events in the energy interval $2.5 \mathrm{\
GeV} \leqslant Q \leqslant 28 \mathrm{\ GeV}$. By fitting the data
on hadron multiplicity in the light quark events at low $Q$, we
get the expression:
\begin{equation}\label{N_light_fit}
N_{l\bar{l}}(Q^2) = 2.07 + 1.11 \  \ln Q + 0.54 \ \ln^2 Q \;.
\end{equation}

Putting $Q_0 = 1$ GeV, we find $\Delta y_b = 1.61$. Taking into
account that the last term in
Eq.~\eqref{delta_N2_bound} is negligible,%
\footnote{Since $E(y) < 0.02$ in the region $0 \leqslant y
\leqslant \Delta y_b -1 = 0.61$.}
we get from \eqref{delta_N2_bound}, \eqref{N_light_fit}:
\begin{equation}\label{delta_N2_b_num}
\delta N_b^{(2)} \geqslant 1.07\;.
\end{equation}
Correspondingly, our prediction accounting the revised data on the
multiplicity in events induced by the light quarks,
\begin{equation}\label{delta_bl_upper_bound}
\delta_{bl}^{\mathrm{QCD}} \leqslant 2 n_b - N_{l\bar{l}}(Y_m +
\Delta y_b) = 3.33 \pm 0.38 \;,
\end{equation}
appears to be lower than our previous result $\delta_{bl} =
3.68$~\cite{Petrov:95}. We used the experimental value
\begin{equation}\label{2n_b}
2n_b = 11.10 \pm 0.18 \;.
\end{equation}
The error in measurements of $N_{l\bar{l}}$ was taken to be $\pm
\, 0.34$. Let us stress that our upper bound
\eqref{delta_bl_upper_bound} is very close to the present
experimental value of $\delta_{bl}^{exp}$~\eqref{delta_bl_exp}.

Now let us derive a lower bound on $\delta_{bl}^{\mathrm{QCD}}$.
To do this, let us start from Eq.~\eqref{delta_N}. It is
convenient to
represent the integral in \eqref{delta_N} as a sum of two terms:%
\footnote{We take into account that the region $-\infty < y < -4$
gives a negligible contribution to $\Delta N_b$.}
\begin{eqnarray}\label{delta_N_bl}
\Delta N_b &=& \int\limits_{-4}^{-1} \! dy \, \hat{N}_g (Y_b - y)
\, \Delta E_b(y) + \int\limits_{-1}^{Y_b} \! dy \, \hat{N}_g (Y_m
- y) \, \Delta E_b(y)
\nonumber \\
&=& \Delta N_b^{(1)} + \Delta N_b^{(2)} \;,
\end{eqnarray}
where $Y_b = \ln (m_b^2/Q_0^2) \simeq 3.16$. Consider the first
term in \eqref{delta_N_bl}. One can check that
\begin{equation}\label{Delta_vs_E_1}
\Delta E(y) < 0.18 \, E(y + 5.8)
\end{equation}
in the region $-4 < y < -1$, that leads to the inequality
\begin{equation}\label{Delta_Nb_1}
\Delta N_b^{(1)} < 0.18 \int\limits_{1.8}^{4.8} \! dy \, \hat{N}_g
(Y_b + 5.8 - y) \, \Delta E_b(y) \;.
\end{equation}

The estimations show that $\hat{N}_g (Y_b + 5.8 - y) < 2 \,
\hat{N}_g (4.8 - y)$ when $y$ varies from 1.8 to 4.8. Thus, we
get:
\begin{equation}\label{Delta_Nb_1_estimate}
\Delta N_b^{(1)} < 0.36 \, \left[ N_{l\bar{l}}(Q = 11 \mathrm{\
GeV}) - N_{l\bar{l}}(Q = 2.5 \mathrm{\ GeV}) \right] = 1.54 \pm
0.17 \;.
\end{equation}

The second term in \eqref{delta_N_bl} can be estimated by using
the inequality
\begin{equation}\label{Delta_vs_E_2}
\Delta E(y) < 0.62 \, E(y + 3.5)
\end{equation}
which is valid in the region  $-1 < y < Y_b$. Then
\begin{equation}\label{Delta_Nb_2_estimate}
\Delta N_b^{(2)} < 0.62 \left[ N_{l\bar{l}}(Q = 28 \mathrm{\ GeV})
- N_{l\bar{l}}(Q = 3.5 \mathrm{\ GeV}) \right] = 4.61 \pm 0.30 \;.
\end{equation}

\begin{figure}[ht]
\begin{center}
\epsfysize=7cm \epsffile{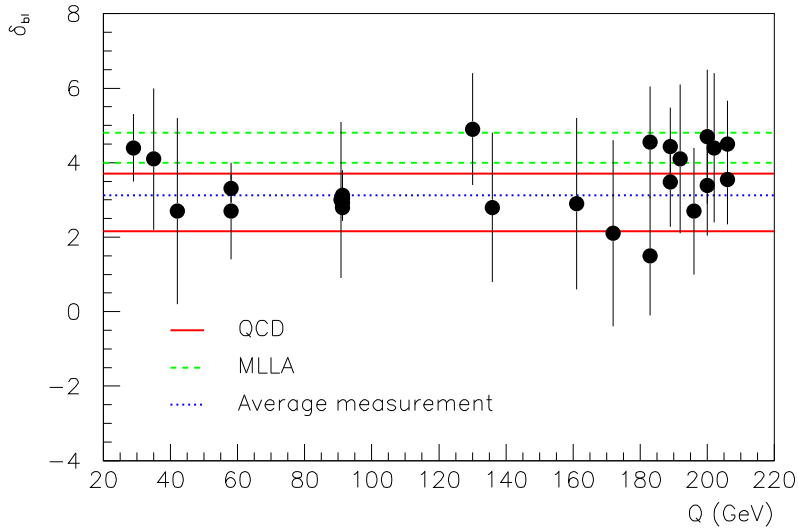}
\end{center}
\caption{Our QCD result for $\delta_{bl}$ (a corridor between the
two solid lines), and the MLLA prediction~\cite{Dokshitzer:06} (a
corridor between the two dashed lines) vs. experimental data.}
\label{fig:data_corr_delta_bl}
\end{figure}

As a result, we obtain from Eqs.~\eqref{delta_Ql_QCD},
\eqref{delta_N} and \eqref{Delta_Nb_1_estimate},
\eqref{Delta_Nb_2_estimate} the lower bound on
$\delta_{bl}^{\mathrm{QCD}}$:
\begin{equation}\label{delta_bl_lower_bound}
\delta_{bl}^{\mathrm{QCD}} > 2.55 \pm 0.39 \;.
\end{equation}
Fig.~\ref{fig:data_corr_delta_bl} demonstrates that our QCD
predictions are very close to the average measurement
$\delta_{bl}^{\rm exp} = 3.12$.

Our results can be compared with the recently published new MLLA
results~\cite{Dokshitzer:06}:
\begin{equation}\label{bl_MLLA}
\delta_{bl}^{\mathrm{MLLA}} = 4.4  \pm 0.4 \;.
\end{equation}
The next-to-MLLA results are:
\begin{align}
\delta_{bl}^{\mathrm{NMLLA}} &= 2.6 \pm 0.4 \label{bl_NMLLA} \;,
\\
\delta_{cl}^{\mathrm{NMLLA}} &= -0.1 \pm 0.4 \label{cl_NMLLA} \;.
\end{align}
Thus, the lowest-order MLLA expression~\eqref{MLLA_prediction} is
not stable against higher order corrections. As it was said above,
the formula which was used in the MLLA approach \eqref{delta_asym}
is only \emph{a part} of the exact QCD formula \eqref{delta_Ql},
\eqref{delta_N} in the approximation $E(y) = \Delta E_Q(y -1) $.
This approximation is quite a rough one (see
Fig.~\ref{fig:deltaE_vs_E}), and the deviation of the function
$\Delta E_Q(y)$ from $\Delta E_Q^{\mathrm{asym}}(y) = y - 3/2$, as
well as the deviation of $E(y)$ from $E^{\mathrm{asym}}(y) = y -
1/2$, cannot be neglected.

Note that the limit $y \rightarrow \infty$ means that the
invariant mass of the gluon jet $k^2$ tends to zero, since $k^2 =
Q^2 \exp(-y)$ (see \eqref{y}). We also think that using the
argument $k_{\perp}^2$ in $N_g$ instead of $k^2$ in the MLLA
scheme (see \cite{Dokshitzer:06} and references therein) has no
reasons. Indeed, $k$ is a time-like vector ($k^2 = k_0^2 -
k_{\perp}^2 - k_{\parallel}^2 > 0$), and the gluon jet with large
transverse momentum $k_{\perp}$ has the small invariant mass
$k^2$. In such a jet, the multiplicity is small, since the phase
space for secondary particles is actually defined by the invariant
mass.



\section{Conclusions}

The formula for the difference between hadron multiplicities in
$e^+e^-$ annihilation into hadrons induced by light and heavy
primary quarks (with $Q$ is a type of a heavy quark) is derived:
\begin{eqnarray}\label{delta_bQ_concl}
\delta_{Ql}^{\mathrm{QCD}} &=& 2n_Q - N_{l\bar{l}}(m_Q^2
\mathrm{e})
\\ \nonumber
&-& \int_{Q_0^2}^{m_Q^2 \mathrm{e}} \! \frac{d k^2}{k^2} \,
\hat{N}_g (k^2) \left[ \Delta E_Q \left( \frac{m_Q^2}{k^2} \right)
- E \left( \frac{m_Q^2 \mathrm{e}}{k^2} \right) \right]
\\ \nonumber
&-& \int_{m_Q^2 \mathrm{e}}^{\infty} \! \frac{d k^2}{k^2} \,
\hat{N}_g (k^2) \, \Delta E_Q \left( \frac{m_Q^2}{k^2} \right) \;.
\end{eqnarray}
Here $\hat{N}_g(k^2)$ describes the average number of charged
hadrons in the gluon jet with the virtuality which varies up to
$k^2$, and $E$, $\Delta E_Q$ are known functions.

By using the data on the hadron multiplicity in light quark events
$N_{l\bar{l}}$, corrected for the detector effects and initial
state radiation effects~\cite{Dokshitzer:06}, we have obtained
from \eqref{delta_bQ_concl} the bounds:
\begin{equation}\label{delta_bl_QCD}
2.2 < \delta_{bl}^{\mathrm{QCD}} < 3.7 \;.
\end{equation}
We would like to emphasize that our estimate does not depend on a
specific choice of the function $N_g(k^2)$, and it is in a good
agreement with the average experimental value $\delta_{bl}^{\rm
exp} = 3.12 \pm 0.14$.

Two last terms in \eqref{delta_bQ_concl}, which are subtracted
from the first one, are positive and numerically large. In
particular, for the case of the beauty ($m_Q = m_b$, $n_Q = n_b$)
the second term in \eqref{delta_bQ_concl} (dominating the third
one) is equal to 1.1.

As a result, the deviation of the MLLA prediction,
\begin{equation}\label{delta_b_MLLA_concl}
\delta_{bl}^{\mathrm{MLLA}} = 2n_b - N_{l\bar{l}}(m_b^2
\mathrm{e}) \;,
\end{equation}
from the exact expression,%
\footnote{This formula is an equivalent compact form of
Eq.~\eqref{delta_bQ_concl} for $Q=b$.}
\begin{equation}\label{delta_bl_concl}
\delta_{bl}^{\mathrm{QCD}} = 2(n_b - n_l) - \int_{Q_0^2}^{m_b^2}
\! \frac{d k^2}{k^2} \, \hat{N}_g (k^2) \, \Delta E_Q \left(
\frac{m_b^2}{k^2} \right) \;,
\end{equation}
appears to be quite significant.

Let us also mention that our numerical predictions for the charm
quark case,
\begin{align}\label{delta_cl_num}
\delta^{\mathrm{QCD}}_{cl} (Q = 91 \mathrm{\ GeV}) &= 1.01 \;,
\nonumber \\
\delta^{\mathrm{QCD}}_{cl} (Q = 170 \mathrm{\ GeV}) &= 0.99 \;,
\end{align}
were derived in \cite{Petrov:95} \emph{before the precise
measurements of $\delta_{cl}$} were
made~\cite{b_c_events_high_energies}.%
\footnote{In the low energy measurements~\cite{Rowson:85,
b_c_events_low_energies}, the total error of $\delta_{cl}$ was
about $\pm 1.5$.}
As one can see, our value \eqref{delta_cl_num} is in a nice
agreement with the average experimental value \eqref{delta_cl_exp}
(see also Fig.~\ref{fig:data_corr_delta_bl}).

Some part of the results of this paper was published in
\cite{Kisselev:07}. We are indebted to W.~Ochs, the correspondence
with whom stimulated, to some extent, the appearance of the present
paper.

The work is partially supported by the grant RFBR-06-02-16031.



\setcounter{equation}{0}
\renewcommand{\theequation}{A.\arabic{equation}}

\section*{Appendix A}
\label{app:A}

Let us consider the average multiplicity of off-shell quanta with
the ``mass'' $Q_0$ in a gluon jet whose invariant mass is $p^2$.
It obeys the following integral equation~\cite{Kisselev:82,
Kisselev:review}:
\begin{eqnarray}\label{gluon_mult}
n_g(p^2, Q^2_0) &=& p^2 \delta (p^2 -  Q^2_0) + \int_{Q^2_0}^{p^2}
\frac{d l^2}{l^2} \int_{0}^{1} d z \, \theta (z p^2 - l^2) \,
\theta (zk^2 - p^2)
\nonumber \\
&\times & \frac{\alpha _s(p^2 z)}{2 \pi } \, \hat P_{gg}(z) \,
n_g(l^2, Q^2_0) \;.
\end{eqnarray}
Here $k^2$ is the virtuality of the parent quark which emits this
gluon jet, and $\hat P_{gg}(z)$ is Altarelli-Parisi time-like
decay function.

The inequality
\begin{equation}\label{kinem_bound}
z \geqslant \frac{l^2}{p^2}
\end{equation}
in Eq.~\eqref{gluon_mult} is a kinematical bound, while the bound
\begin{equation}\label{dynam_bound}
z \geqslant \frac{p^2}{k^2}
\end{equation}
is a dynamical one. The latter is nothing but the angle ordering
rewritten in terms of momentum fractions and virtualities (see
Refs.~\cite{Kisselev:82}, \cite{Kisselev:review}).%
\footnote{Namely, the emission angle of the secondary gluon with
the virtuality $l^2$ is less than the angle of the primary gluon
emission off the parent quark.}

Assuming that parton shower develops mainly via \emph{soft}
gluons, one can put in \eqref{gluon_mult}
{\allowdisplaybreaks
\begin{equation}\label{P_gg}
\hat  P_{gg}(z) \Big|_{z \ll 1} \simeq 2 C_A \, \frac{1}{z} \;.
\end{equation}
Then the sequence of equations~\cite{Kisselev:thesis},
\begin{eqnarray}\label{eq_N_g}
\lefteqn{N_g(k^2, Q^2_0) - 1}
\nonumber \\
&=&  C_A  \int_{Q^2_0}^{k^2} \frac{d p^2}{p^2}  \int_{
Q^2_0}^{p^2} \frac{d l^2}{l^2} \int_{0}^{1} \frac{d z}{z} \theta
(z  p^2 - l^2) \theta (z k^2 - p^2) \frac{\alpha_s(z p^2)}{\pi}
n_g(l^2, Q^2_0)
\nonumber \\
&=& C_A \int_{Q^2_0}^{k^2} \frac{d p^2}{p^2} \theta (\sqrt{k^2
Q^2_0} - p^2)   \int_{Q^2_0}^{p^2} \frac{d l^2}{l^2} \int_{0}^{1}
\frac{d z}{z} \theta (z p^2 - l^2) \frac{\alpha_s(z p^2)}{\pi}
n_g(l^2, Q^2_0)
\nonumber \\
&+&  C_A  \int_{Q^2_0}^{k^2}  \frac{d  p^2}{p^2} \theta (p^2 -
\sqrt{k^2 Q^2_0}) \left[  \int_{Q^2_0}^{p^2} \frac{d l^2}{l^2}
\theta (p^4 - k^2 l^2) \int_{0}^{1} \frac{d z}{z} \theta (z k^2 -
p^2) \right.
\nonumber \\
&+& \left. \int_{Q^2_0}^{p^2} \frac{d l^2}{l^2} \theta (k^2 l^2 -
p^4) \int_{0}^{1} \frac{d z}{z} \theta (z p^2 - l^2) \right]
\frac{\alpha_s(z p^2)}{\pi} n_g(l^2, Q^2_0)
\nonumber \\
&=& C_A \int_{Q^2_0}^{k^2} \frac{d p^2}{p^2} \theta (\sqrt{k^2
Q^2_0} - p^2) \int_{Q^2_0}^{p^2} \frac{d r^2}{r^2}
\frac{\alpha_s(r^2)}{\pi} N_g(r^2, Q^2_0)
\nonumber \\
&+&  C_A  \int_{Q^2_0}^{k^2}  \frac{d p^2}{p^2} \theta (p^2 -
\sqrt{k^2 Q^2_0})  \left[ \int_{Q^2_0}^{p^2} \frac{d r^2}{r^2}
\theta (r^2  k^2 - p^4) \int_{ Q^2_0}^{r^2} \frac{d l^2}{l^2}
\theta (p^4 - k^2 l^2) \right.
\nonumber \\
&+& \left. \int_{Q^2_0}^{p^2} \frac{d r^2}{r^2}  \theta (r^2 k^2 -
p^4) \int_{Q^2_0}^{r^2} \frac{d l^2}{l^2} \theta (k^2 l^2 - p^4)
\right] \frac{\alpha_s(r^2)}{\pi} n_g(l^2, Q^2_0)
\nonumber \\
&=& C_A \int_{Q^2_0}^{k^2} \frac{d p^2}{p^2} \theta (\sqrt{k^2
Q^2_0} - p^2)   \int_{Q^2_0}^{p^2} \frac{dr^2}{r^2}
\frac{\alpha_s(r^2)}{\pi} N_g(r^2, Q^2_0)
\nonumber \\
&+&  C_A  \int_{Q^2_0}^{k^2}  \frac{d  p^2}{p^2} \theta (p^2 -
\sqrt{k^2  Q^2_0}) \int_{Q^2_0}^{p^2}  \frac{d r^2}{r^2} \theta
(k^2  r^2 - p^4)  \frac{\alpha_s(r^2)}{\pi} N_g(r^2, Q^2_0)
\nonumber \\
&=&  C_A  \int_{Q^2_0}^{k^2} \frac{d  p^2}{p^2} \int_{Q^2_0}^{p^2}
\frac{d r^2}{r^2} \frac{\alpha_s(r^2)}{\pi} N_g(r^2, Q^2_0)
\nonumber \\
&-&  C_A  \int_{Q^2_0}^{k^2} \frac{d  p^2}{p^2} \theta (p^2 -
\sqrt{k^2  Q^2_0}) \int_{Q^2_0}^{p^2}  \frac{d r^2}{r^2} \theta
(p^4 - k^2  r^2) \frac{\alpha_s(r^2)}{\pi} N_g(r^2, Q^2_0)
\nonumber \\
&=&  C_A  \int_{Q^2_0}^{k^2} \frac{d p^2}{p^2} \int_{Q^2_0}^{p^2}
\frac{d r^2}{r^2} \frac{\alpha_s(r^2)}{2 \pi} N_g(r^2, Q^2_0) \;,
\end{eqnarray}
}
results in the following formula for $N_g(k^2, Q^2_0)$:
\begin{equation}\label{eq_N_g_final}
N_g(k^2, Q^2_0) = 1 +  C_A \int_{Q^2_0}^{k^2} \frac{d p^2}{p^2}
\int_{Q^2_0}^{p^2} \frac{d r^2}{r^2} \frac{\alpha_s(r^2)}{2\pi}
N_g(r^2, Q^2_0) \;.
\end{equation}

From \eqref{eq_N_g_final} we obtain the differential equation:
\begin{equation}\label{diff_eq_N_g}
\left(k^2 \frac{d}{d k^2} \right)^2 N_g(k^2, Q^2_0) = C_A
\frac{\alpha _s(k^2)}{2 \pi} N_g(k^2, Q^2_0) \;,
\end{equation}
with the boundary conditions
\begin{equation}\label{boundary_cond}
N_g(k^2, Q^2_0) \Big|_{k^2 = Q^2_0}  =  1, \quad  k^2 \frac{d}{d
k^2} N_g(k^2, Q^2_0) \Big|_{k^2 = Q^2_0} = 0 \;.
\end{equation}

This equation has the solution:
\begin{eqnarray}\label{solution_N_g}
N_g(k^2, Q^2_0) &=& \sqrt{\frac{2 C_A}{\pi b} \ln
\frac{k^2}{\Lambda ^2}} \left[ K_1 \left( \sqrt{\frac{2 C_A}{\pi
b} \ln \frac{k^2}{\Lambda^2}} \right) I_0 \left(\sqrt{\frac{2
C_A}{\pi b} \ln \frac{ Q^2_0}{\Lambda^2}} \right) \right.
\nonumber \\
&+& \left. I_1 \left(\sqrt{\frac{2 C_A}{\pi  b} \ln
\frac{k^2}{\Lambda ^2}} \right)  K_0  \left( \sqrt{\frac{2
C_A}{\pi b} \ln \frac{ Q^2_0}{\Lambda^2}} \right) \right] \;,
\end{eqnarray}
with the asymptotics
\begin{equation}\label{asym_N_g}
\left. N_g(k^2, Q^2_0) \right|_{k^2 \gg Q^2_0} \simeq \exp
\left(\sqrt{\frac{2 C_A}{\pi b} \ln \frac{k^2}{\Lambda^2}} \right)
\;.
\end{equation}
Here $b = (33 - 2 N_f)/12 \pi$, where $N_f$ is a number of
flavors.

Let us stress that the equation for \emph{an isolated} gluon jet
(which has no parent parton with virtuality $k^2$) would be of the
form:
\begin{eqnarray}\label{gluon_mult_isolated}
n_g^{\mathrm{isol}}(p^2, Q^2_0) &=& p^2 \delta (p^2 - Q^2_0) +
\int_{Q^2_0}^{p^2} \frac{d l^2}{l^2} \int_{0}^{1} dz \, \theta (z
p^2 - l^2)
\nonumber \\
&\times & \frac{\alpha_s(p^2 z)}{2 \pi } \hat P_{gg}(z) \,
n_g(l^2, Q^2_0) \;,
\end{eqnarray}
that leads to the formula:
\begin{equation}\label{eq_N_g_no_ordering}
N_g(k^2, Q^2_0) = 1 +  2 C_A  \int_{Q^2_0}^{k^2} \frac{d p^2}{p^2}
\int_{Q^2_0}^{p^2} \frac{d r^2}{r^2} \frac{\alpha_s(r^2)}{2 \pi}
N_g(r^2, Q^2_0).
\end{equation}
This equation results in a wrong expression which does not take
into account the interference effects:
\begin{equation}\label{asym_N_g_no_ordering}
\left. N_g(k^2, Q^2_0) \right|_{k^2 \gg Q^2_0} \simeq \exp
\left(2\sqrt{\frac{C_A}{\pi b} \ln \frac{k^2}{\Lambda^2}} \right)
\;.
\end{equation}

\setcounter{equation}{0}
\renewcommand{\theequation}{B.\arabic{equation}}

\section*{Appendix B}
\label{app:B}

Now we will reproduce the asymptotic relation between the average
multiplicity in the light quark event and that in the gluon jet by
using the Altarelli-Parisi decay functions. Let $l$ be a
4-momentum of the primary quark which emits a massive gluon jet.
The ladder diagram in Fig.~\ref{fig:ladder} leads to the equation
\begin{align}\label{qq_mult_full}
N_{l \bar{l}}(Q^2,Q_0^2) \Big|_{Q^2 \gg Q_0^2} &=
\int_{Q_0^2}^{Q^2} \frac{dl^2}{l^2} \int_{Q_0^2/l^2}^{1} \!\! d z
\, \frac{\alpha_s(z l^2)}{2\pi} \int_{Q_0^2}^{z l^2}
\frac{dk^2}{k^2}
\nonumber \\
& \times \hat{P}_{qg} \left( z, \frac{k^2}{l^2} \right) \,
n_g(k^2, Q_0^2)
\end{align}
(for simplicity, here and below we omit the contribution from the
leading hadrons, $2n_q$).

In the leading logarithm approximation,
\begin{equation}\label{AP_function}
\hat{P}_{qg} \left( z, \frac{k^2}{l^2} \right) \simeq 2 C_F \,
\frac{1}{z} \;,
\end{equation}
one comes to the expression ($r^2 = z l^2$):
\begin{align}\label{qq_mult_appr}
N_{l \bar{l}}(Q^2, Q_0^2) &= C_F \int_{Q_0^2}^{Q^2}
\frac{dl^2}{l^2} \int_{Q_0^2}^{l^2} \frac{dr^2}{r^2}
\frac{\alpha_s(r^2)}{\pi} \, N_g(r^2, Q_0^2)
\nonumber \\
&= C_F \int_{Q_0^2}^{Q^2} \frac{dr^2}{r^2}
\frac{\alpha_s(r^2)}{\pi} \, N_g(r^2, Q_0^2) \int_{r^2}^{Q^2}
\frac{dl^2}{l^2}
\nonumber \\
&= C_F \int_{Q_0^2}^{Q^2} \frac{dr^2}{r^2}
\frac{\alpha_s(r^2)}{\pi} \, \ln \frac{Q^2}{r^2} \, N_g(r^2,
Q_0^2) \;,
\end{align}
where we have used the relation:
\begin{equation}\label{N_g_vs_n_g_off-shell}
N_g(r^2, Q_0^2) = \int^{r^2}_{Q_0^2} \!\! \frac{d k^2}{p^2} \,
n_g(k^2, Q_0^2) \;.
\end{equation}
The integral equation for $N_g(r^2, Q_0^2)$ has been obtained in
Appendix~A (see Eq.~\eqref{eq_N_g_final}).

The formula \eqref{qq_mult_appr} can be represented as%
\footnote{We have added the non-leading term -1/2 to
$\ln(Q^2/r^2)$ in deriving \eqref{qq_mult_asym} from
\eqref{qq_mult_appr}.}
\begin{equation}\label{qq_mult_asym}
N_{l \bar{l}}(Q^2, Q_0^2) = C_F \!\int_{Q_0^2}^{Q^2} \!
\frac{dk^2}{k^2} \,\frac{\alpha_s(k^2)}{\pi} \,
E^{\mathrm{asym}}(Q^2, k^2) \, N_g(k^2, Q_0^2) \;,
\end{equation}
with the function $E^{\mathrm{asym}}(Q^2, k^2)$ defined above
\eqref{E_asym}. It is worth to compare this approximate expression
with our exact formula \eqref{qq_mult}.

At large $Q^2$, we obtain from \eqref{qq_mult_appr} and
\eqref{asym_N_g} the well-known \emph{asymptotic} relation:
\begin{equation}\label{e+e-mult_vs_N_g}
N_{l \bar{l}}(Q^2,Q_0^2) \Big|_{Q^2 \gg Q_0^2} \simeq
\frac{2C_F}{C_A} \, N_g(Q^2, Q_0^2) \;.
\end{equation}
Remember that $N_g(Q^2, Q_0^2)$ describes the average number of
virtual partons in the gluon jet whose invariant mass \emph{varies
from} $Q_0$ \emph{up to} $Q$.

Our formula \eqref{qq_mult} should be reproduced from the starting
equation \eqref{qq_mult_full} provided:
\begin{enumerate}
    \item
    Contributions from both the ladder (Fig.~\ref{fig:ladder})
    \emph{and interference} diagrams (Fig.~\ref{fig:cross}) are added
    together in deriving the expression for $N_{l \bar{l}}(Q^2, Q_0^2)$
    \item
    Both non-singular terms in variable $z$ and power corrections
    $\mathrm{O}(k^2/l^2)$ are taken into account in
    $\hat{P}_{qg}(z, k^2/l^2)$
\end{enumerate}

\setcounter{equation}{0}
\renewcommand{\theequation}{C.\arabic{equation}}

\section*{Appendix C}
\label{app:C}

The average number of hadrons in $e^+e^-$ annihilation is, of
course, gauge invariant quantity. However, in perturbative QCD we
calculate the multiplicity of \emph{virtual partons}.
\emph{Apriori} one can not be sure that it does not depend on a
gauge.

Unfortunately, this important problem was not studied to a
considerable extent by other authors. As a few exceptions, let us
mention Refs.~\cite{Furmanski:79} and \cite{Bassetto:82}. In the
first paper the gauge invariance of the multiplicity in $e^+e^-$
annihilation has been checked in one-loop approximation. In the
second one the gauge dependence was considered for a fixed
coupling constant and without accounting for interference effects.

Below we will analyze a possible gauge dependence of the partonic
multiplicity in the case of light primary quark in a class of
axial gauges. It is given by Eq.~\eqref{qq_mult_full} in the gauge
$n_{\mu} A^{\mu} = 0$ with the gauge vector
\begin{equation}\label{gauge_quark}
n_{\mu} = \frac{1}{\sqrt{2}}(1, 0, -1)
\end{equation}
($z$-axis is chosen along a 3-momentum of a primary quark in the
c.m.s. of colliding leptons). The argument of the decay function
$\hat{P}_{qg} (z)$ in \eqref{qq_mult_full} is the ratio
\begin{equation}\label{z}
z = \frac{kn}{ln} \;.
\end{equation}
The emission of the gluon jets from the primary antiquark is
suppressed in this gauge \eqref{gauge_quark}.

Analogously, in the gauge
\begin{equation}\label{gauge_antiquark}
n_{\mu} = \frac{1}{\sqrt{2}}(1, 0, 1)
\end{equation}
the massive gluon jet is emitted by the antiquark while its
emission from the quark is suppressed.

Let us choose the gauge in which both quark and antiquark make
comparable contributions to the emission:
\begin{equation}\label{Hamilton_gauge}
n_{\mu} = (1, 0, 0) \;.
\end{equation}
Then Eq.~\eqref{qq_mult_full} is modified as follows:
\begin{align}\label{mult_quark_antiquark}
N_{l \bar{l}}(Q^2, Q_0^2) &= \int_{Q_0^2}^{Q^2} \frac{dl^2}{l^2}
\int_{Q_0^2/l^2}^{1} \!\! d z \, \frac{\alpha_s(z l^2)}{2\pi} \,
\int_{Q_0^2}^{z l^2} \frac{dk^2}{k^2} \hat{P}_{qg}(z) \, n_g(k^2,
Q_0^2)
\nonumber \\
&+ \int_{Q_0^2}^{Q^2} \frac{dl^2}{l^2} \int_{Q_0^2/l^2}^{1} \!\!
dz \, \frac{\alpha_s(z l^2)}{2\pi} \, \int_{Q_0^2}^{z l^2}
\frac{dk^2}{k^2} \hat{P}_{\bar{q}g} (z) \, n_g(k^2, Q_0^2) \;,
\end{align}
where
\begin{equation}\label{AP_quark_antiquark}
\hat P_{qg}(z) =  \hat{P}_{\bar{q}g} (z) \simeq 2 C_A \,
\frac{1}{z +
\begin{displaystyle} \frac{l^2}{Q^2} \end{displaystyle}}
\end{equation}
at small $z$. These relations mean that the massive jets are
emitted by the quark and antiquark with the same probability in
the gauge~\eqref{Hamilton_gauge}.

By omitting non-leading terms, we get from
\eqref{mult_quark_antiquark}, \eqref{AP_quark_antiquark} the
expression,
\begin{align}\label{mult_quark_antiquark_appr}
N_{l \bar{l}}(Q^2, Q_0^2) &= 2C_F \int_{Q_0^2}^{Q^2}
\frac{dl^2}{l^2} \int_{Q_0^2}^{l^2} \frac{dr^2}{r^2 +
\begin{displaystyle} \frac{l^4}{Q^2} \end{displaystyle}}
\frac{\alpha_s(r^2)}{\pi} \, N_g(r^2, Q_0^2)
\nonumber \\
&= 2C_F \int_{Q_0^2}^{Q^2} dr^2 \, \frac{\alpha_s(r^2)}{\pi} \,
N_g(r^2, Q_0^2) \int_{r^2}^{Q^2} \frac{dl^2}{l^2} \frac{1}{r^2 +
\begin{displaystyle} \frac{l^4}{Q^2} \end{displaystyle}}
\nonumber \\
&= C_F \int_{Q_0^2}^{Q^2} \frac{dr^2}{r^2}
\frac{\alpha_s(r^2)}{\pi} \, \ln \frac{Q^2}{r^2} \, N_g(r^2,
Q_0^2) \;,
\end{align}
which coincides with formula~\eqref{qq_mult_appr}.

Note that in the gauge~\eqref{Hamilton_gauge} the decay function
in evolution equation~\eqref{gluon_mult} is of the form:
\begin{equation}\label{AP_quark}
\hat P_{gg}(z) \simeq 2C_A \, \frac{1}{z + \begin{displaystyle}
\frac{p^2}{Q^2} \end{displaystyle}} \;.
\end{equation}
This results in the effective cut on the integration variable $z$
from below:
\begin{equation}\label{gauge_bound}
z \geqslant \frac{p^2}{Q^2} \;.
\end{equation}
As one can see from \eqref{dynam_bound} and \eqref{gauge_bound},
it is the dynamical bound \eqref{dynam_bound} that smoothes out
the singularity of $\hat P_{gg}(z)$ at $z=0$, but not the bound
\eqref{gauge_bound} arising from the gauge vector. The latter can
be safely omitted in Eq.~\eqref{gluon_mult}.

Thus, we conclude that both the relation of the light quark
multiplicity $N_{l \bar{l}}(k^2,Q_0^2)$ with the gluon
multiplicity $N_g(Q^2, Q_0^2)$, and the evolution equation for
$N_g(k^2, Q_0^2)$ do not depend on the gauge vector $n_{\mu}$.
This conclusion is also valid for a general case, $n_{\mu} = (n_0,
0, n_{\shortparallel})$, where $n_0 \neq \pm
n_{\shortparallel}$~\cite{Kisselev:82,Kisselev:review}. Note that
the proof of the gauge invariance needs an account of the
destructive interference in the emission of the gluon jets, that
leads to the condition \eqref{dynam_bound}.




\end{document}